\pdfoutput=1
\documentclass[sigconf,nonacm,natbib=false]{acmart}
\usepackage[utf8]{inputenc}
\usepackage[T2A,T1]{fontenc}
\usepackage{palatino}
\usepackage{amsmath,amsfonts}
\usepackage{hyperref}
\usepackage{cleveref}
\usepackage{svg}
\usepackage{url}
\usepackage{stackengine}
\usepackage{csquotes}
\usepackage[style=ACM-Reference-Format,backend=bibtex,sorting=none]{biblatex}
\usepackage{xurl}
\usepackage{fancyhdr}

\fancypagestyle{firststyle}
{
   \fancyhf{}
   \chead{\raisebox{2em}[0pt][0pt]{\textit{--------------- In 1\textsuperscript{st} ACM Workshop on Software Supply Chain Offensive Research and Ecosystem Defenses ---------------}}}
}

\title[Talking Trojan: Analyzing an Industry-Wide Disclosure]{Talking Trojan:\\Analyzing an Industry-Wide Disclosure\vspace{0.5em}}

\author{Nicholas Boucher}
\orcid{0000-0002-5674-3730}
\affiliation{%
  \institution{University of Cambridge}
  \country{}
}
\email{nicholas.boucher@cl.cam.ac.uk}
\email{}

\author{Ross Anderson}
\orcid{0000-0001-8697-5682}
\affiliation{%
  \institution{Universities of Cambridge \& Edinburgh}
  \country{}
}
\email{ross.anderson@cl.cam.ac.uk}
\email{}

\date{}

\hypersetup{
    pdftitle={Talking Trojan: Analyzing an Industry-Wide Disclosure},
    pdfauthor={Boucher and Anderson}
}

\def\ucr{\stackinset{c}{}{c}{-.2pt}{%
  \textcolor{white}{\sffamily\bfseries\small ?}}{%
  \rotatebox{45}{$\blacksquare$}}}

\newcommand{\cyrillicp}{{\fontencoding{T2A}\selectfont р}}

\begin{document}

\begin{abstract}
While vulnerability research often focuses on technical findings and post-public release industrial response, we provide an analysis of the rest of the story: the coordinated disclosure process from discovery through public release. The industry-wide `Trojan Source' vulnerability which affected most compilers, interpreters, code editors, and code repositories provided an interesting natural experiment, enabling us to compare responses by firms versus nonprofits and by firms that managed their own response versus firms that outsourced it. We document the interaction with bug bounty programs, government disclosure assistance, academic peer review, and press coverage, among other topics. We compare the response to an attack on source code with the response to a comparable attack on NLP systems employing machine-learning techniques. We conclude with recommendations to improve the global coordinated disclosure system.
\end{abstract}

\maketitle
\thispagestyle{firststyle}

\section{Introduction}

Following the discovery of a vulnerability that affected most languages, code editors and repositories, we began an industry-wide coordinated disclosure. The vulnerability, the Trojan Source attack, enables the code seen by a human reviewer to differ from that seen by a compiler or interpreter~\cite{boucher_trojansource_2021}. This paper describes the real-world experience of trying to get this vulnerability fixed.

As the global software supply chain continues to increase in complexity, the ability of software maintainers to respond effectively to vulnerabilities with broad impact becomes ever more important. In 2021 we had two high-profile examples: the Log4j vulnerability which enabled easy remote code execution across a significant portion of the global production ecosystem~\cite{log4j_2021}, and the SolarWinds incident which allowed APT backdoor access to the production systems of thousands of firms and dozens of government departments~\cite{solarwinds_2021}. Improving the ability of the industry to respond to threats quickly and cohesively across organizations is so important to the security of critical national infrastructure that the US White House has called attention to the protection of the software supply chain~\cite{biden_2021}.

The exceptionally broad scope of the Trojan Source vulnerability meant that its disclosure provided a natural experiment to explore the global coordinated disclosure system, the reporting tools currently available to security researchers, the financial incentives of bounty programs, the process for security engagement with the open source community, and the challenging interplay of large-scale coordinated disclosure with peer-reviewed publication.

In this paper, we make the following contributions:
\begin{itemize}
    \item We present a real-world evaluation of the coordinated disclosure system via a unique case study.
    \item We empirically compare the responses to disclosures from subgroups within the industry.
    \item We document which methods and tools succeeded or failed in the disclosure process.
    \item We recommend specific, tactical changes to improve the global coordinated disclosure system.
\end{itemize}
\section{The Vulnerability}

\begin{figure*}[t]
  \centering
  \begin{minipage}[b]{0.49\textwidth}
    \centering
    \frame{\includegraphics[width=\linewidth]{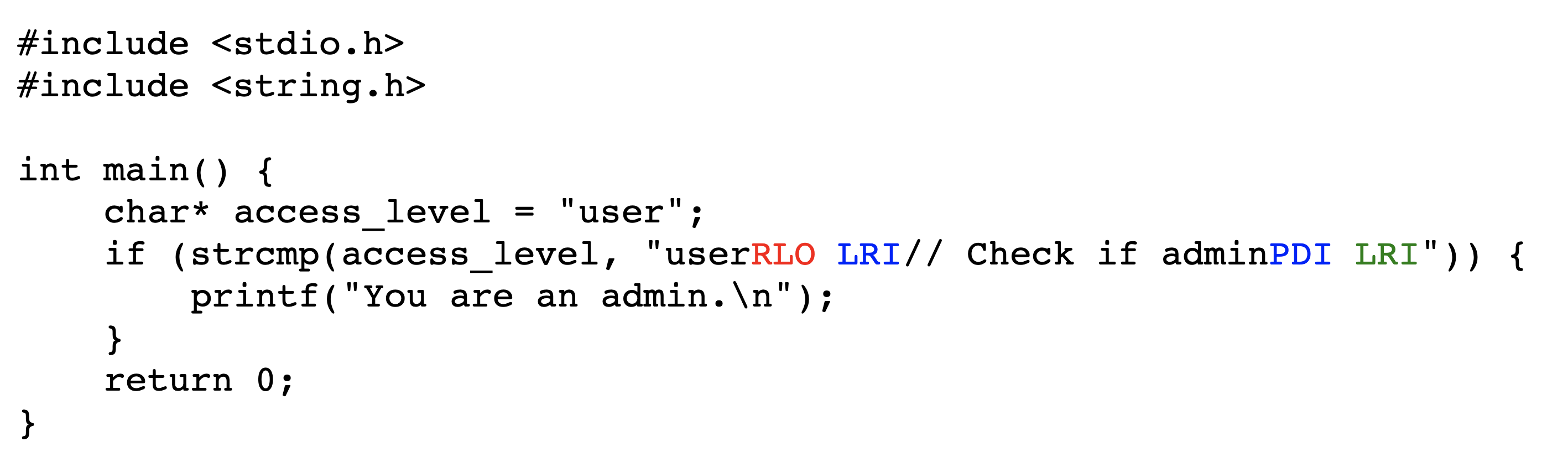}}
    \caption{Example of a Trojan-Source Attack in C where control character are visualized in color~\cite{boucher_trojansource_2021}.}
    \label{fig:c-example-encoded}
  \end{minipage}
  \hfill
  \begin{minipage}[b]{0.49\textwidth}
    \centering
    \frame{\includegraphics[width=\linewidth]{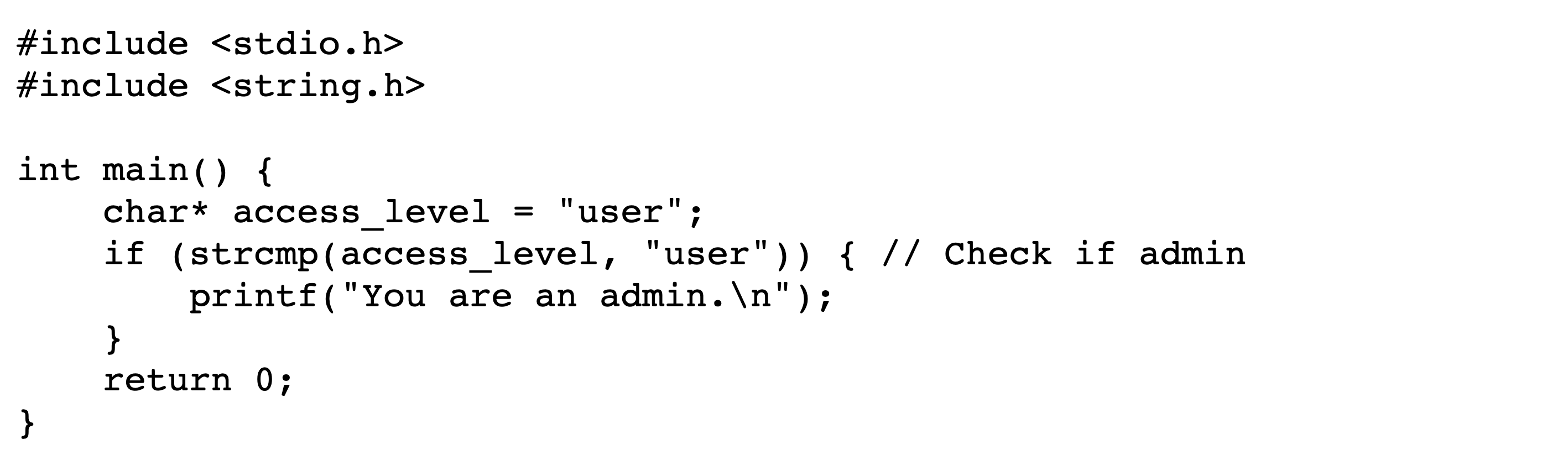}}
    \caption{Example of a Trojan-Source Attack in C as rendered to users~\cite{boucher_trojansource_2021}.}
    \label{fig:c-example-rendered}
  \end{minipage}
\end{figure*}

In this section, we will briefly describe the Trojan Source vulnerability. For a more detailed presentation, we refer to the technical paper describing the attack~\cite{boucher_trojansource_2021}.

Unicode is a text-encoding standard aiming to capture all global character sets into a single specification~\cite{unicode_2021}. Most text, whether in documents, websites, or source code files, is now represented as Unicode.

As a consequence of its broad linguistic support, Unicode supports a wide range of control characters and special symbols. In particular, it supports both left-to-right and right-to-left text where the display order is set by the standardized Bidirectional Algorithm~\cite{unicode_bidi_2021}. The specification also defines control characters that allow fine-grained control of text display directionality. This means that the logically encoded order of text and its visual display order can differ, in ways that adversaries can exploit. Such attacks had already been discussed in the context of natural-language processing (NLP) systems based on machine learning (ML)~\cite{boucher_2022_badchars}.

In the Trojan Source attack, adversaries use bidirectional (`bidi') control characters to modify the display order of characters in source code, so that the logic seen by human code reviewers differs from the logic seen by the compiler or interpreter. In a nutshell, adversaries use bidi control characters to anagram adversarial source code so that it appears to be benign. 

To smuggle malicious control characters into source code, adversaries can place them into comments and string literals. The key insight is that the reordered characters can display statements with valid syntax in order to deceive code reviewers. An example is given in \Cref{fig:c-example-encoded}, where the underlying source code encoding as seen by the compiler is visualized such that  \texttt{RLO} sets right-to-left,  \texttt{LRI} sets left-to-right, and \texttt{PDI} resets directionality. \Cref{fig:c-example-rendered} shows how this same code is deceptively rendered to human viewers.

There are two further variants of the attack. The second version uses homoglyphs, characters that look similar but are distinct. If an adversary defines a Trojan function in an upstream dependency that uses homoglyphs to appear visually identical to another function, they can implement adversarial logic in that function and then deceptively call the Trojan function in a downstream project.

For example, imagine defining a function called \texttt{print}, but where the \texttt{p} is actually the Cyrillic character \texttt{\cyrillicp}\footnote{Unicode character \texttt{U+440}}. This new function \texttt{\cyrillicp rint} could be defined to implement adversarial behavior. If this Trojan function is defined in a package that gets imported into a program's global namespace, an adversary need only make a pull request on the downstream code that references the Trojan function name.

The third and final version of this attack uses invisible characters -- characters that render to the absence of a glyph -- to spoof multiline comments. Consider C-style comments which start with \texttt{/*} and are terminated with \texttt{*/}. An adversary can place an invisible character such as the Zero-Width Space (ZWSP)\footnote{Unicode character \texttt{U+200B}} between the \texttt{*} and \texttt{/} character to prevent a compiler from terminating a multiline comment. Thus, although a comment might appear to  a code reviewer to terminate, it actually extends across several lines until the next unperturbed \texttt{*/} sequence.

These techniques all enable an adversary to write vulnerabilities into source code files at the encoding level. These vulnerabilities are invisible to humans viewing source code, unless they are using a tool that will somehow alert them, or block the techniques. The ability to hide vulnerabilities in plain sight is particularly valuable for supply chain attacks. Many critical systems depend on components maintained in open-source projects, to which anyone can contribute, and where the main defense against malicious code contribution is human source code review.
\section{Disclosure}

\begin{figure}[t]
    \centering
    \caption{Trojan Source discovery, disclosure, and release timeline}
    \vspace{1em}
    \includesvg[width=.8\columnwidth]{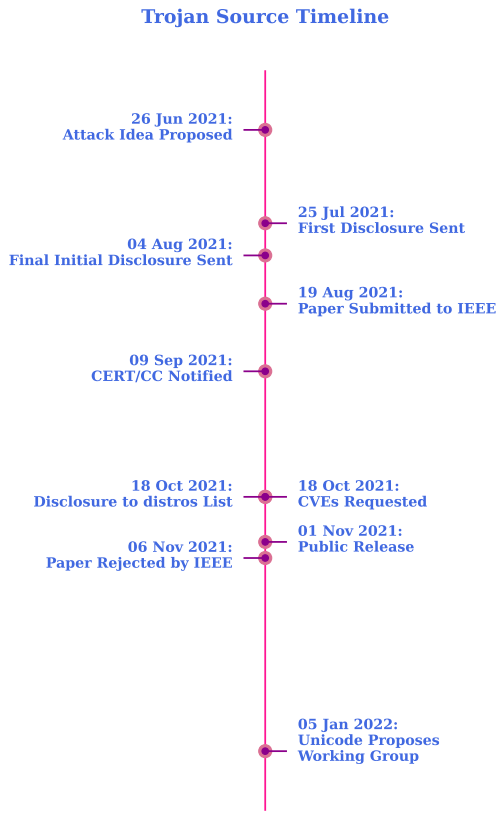}
    \label{fig:timeline}
\end{figure}

A broad timeline of the Trojan Source coordinated disclosure process is shown in \Cref{fig:timeline}. We will now walk through the process in more detail.

\subsection{Initial Disclosures}

We first identified Trojan Source attacks on June 26, 2021, largely building on previous work in adversarial natural language processing~\cite{boucher_2022_badchars}, which we adapted to compilers. After implementing a series of proofs of concept, we found that our attack pattern worked against almost every modern language we tested, including C, C++, C\#, JavaScript, Java, Rust, Go, and Python. We also discovered that the attacks did not trigger any visual alarms in the most common code editors or in the web frontends to online code repositories. Any combination of a vulnerable language and a vulnerable editor or viewer could potentially allow an exploit.

We felt obliged to notify the owners or maintainers of each product in which we observed the vulnerability. We therefore wrote a two-page summary of the attack, including a variety of mitigation techniques, and sent it to 13 companies and open-source organizations over the 11-day period between July 25th and August 4th.

\begin{figure}[t]
    \centering
    \includesvg[width=\columnwidth]{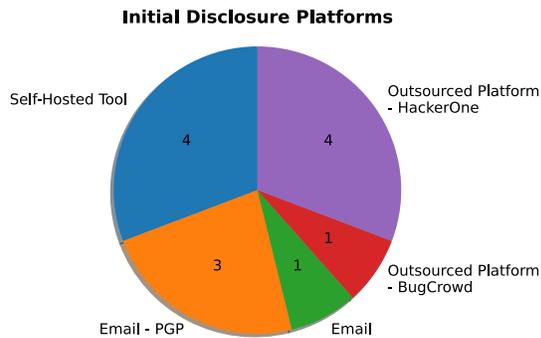}
    \caption{The disclosure platforms used for the initial set of disclosures}
    \label{fig:platforms}
\end{figure}

The recipients used a variety of different platforms for receiving disclosures, which we illustrate in \Cref{fig:platforms}. The disclosure platforms were divided between five outsourced platforms and eight self-hosted tools, of which four involved a web form, three asked for PGP-encrypted email and one requested plaintext email.

\subsection{Outsourced Platforms}
\label{sec:outsourced_platforms}

Of the five initial recipients who used an outsourced platform, four used HackerOne~\cite{hackerone} and one used BugCrowd~\cite{bugcrowd}. These platforms' business model is to collect incoming vulnerability reports and triage them according to an agreed scope before sending them to the client company. They also handle the mechanics of paying bug bounties, and companies that want one of their systems tested can use them to advertise bounties to security researchers who work with their platform.

Our experience with these platforms was mixed. Initial responses to disclosures tended to be fast, often resulting in a reply within a few hours. However, the quality of responses tended to be low, with many reports closed quickly as non-threats.

We learned that these platforms focus on the identification of well-known vulnerability patterns such as buffer overflows and cross-site scripting that are easily demonstrated. However, they perform poorly with novel threats that do not fit the usual patterns. One engineer later remarked to us that the platforms operate according to scopes defined by their customers, and that defining a scope for vulnerability reporting can be hard. As a result, novel vulnerabilities are likely to be dismissed.

We found one way past this problem: to request on the platform's discussion board that the disclosure be reviewed by a full-time employee of the client company. This usually cut through, and once our reports were reviewed by client company staff they were typically identified as relevant. This phenomenon was not unique to outsourced platforms, though -- on multiple occasions we found that disclosures to companies who hosted their own reporting tools were stalled or ignored. Our strategy then was to reach out to pre-existing contacts in the affected company and ask them to look at the case. This would usually result in progress. Presumably some firms that run their disclosure systems internally also have scope restrictions for their first responders.

\subsection{Bug Bounties}

Many companies have bug bounty programs that offer money for the embargoed disclosure of vulnerabilities. We noticed a correlation between having a bug bounty program and using an outsourced disclosure platform (r=0.65, n=19). Another strong indicator of whether a bug bounty would ultimately be paid was whether the receiving organization was a commercial firm rather than a nonprofit open-source project (r=0.46, n=19).

Of the 13 organizations to which we made an initial disclosure, nine had bug bounty programs. Of these, five paid bounties in the amounts of \$1,337, \$525, \$1,370, \$5,000, and \$3,000 USD, totaling \$11,232.

\begin{figure}[t]
    \centering
    \includesvg[width=\columnwidth]{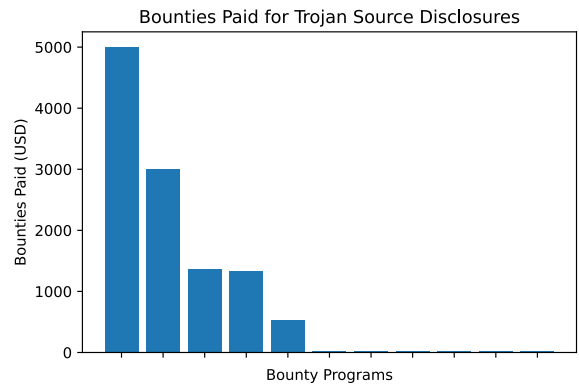}
    \caption{Amounts paid by each bug bounty program submission}
    \label{fig:bounties}
\end{figure}

After we sent the initial round of disclosures, the known impact grew and we sent additional disclosures to other organizations. Two of the new recipients had bounty programs, but neither of them ultimately paid anything. We graph the amounts paid for bounty submissions in \Cref{fig:bounties}.

Two of the five organizations that ultimately paid had initially declined a payment. We received multiple messages in response to our disclosures stating that the disclosures didn't align with the recipient's bounty payment program. This is understandable in terms of what we learned about internal scoping. In two of these cases, recipient company staff eventually agreed a modest payment. 

\subsection{CERT/CC}

The US CERT Coordination Center (CERT/CC) is CISA-backed, CMU-housed institute which provides support for coordinated disclosures~\cite{certcc}. Security researchers can request the assistance of CERT/CC for circulating broad, embargoed disclosures across an affected ecosystem.

We asked for assistance with the coordinated disclosure of Trojan Source attacks from CERT/CC on September 9, 2021. It was accepted on the same day, giving us access to a tool called VINCE, a shared message board that can be used for cross-organization communication. It also provided a central location for us to upload vulnerability descriptions, proofs-of-concept, and vulnerability identifiers.

VINCE provides a platform through which affected vendors can communicate directly with each other, and had been requested by some of the disclosure recipients for coordinating mitigation efforts. It was also helpful to us, as it enabled us to monitor a single location rather than tracking a growing number of email threads and web-based tools. Even with the thirteen disclosures sent in our initial outreach, responding to questions and tracking discussion threads quickly became a multi-week, full-time job.

CERT/CC also added additional vendors to the VINCE case, bringing our total number of advance disclosures to 19.

One downside of using CERT/CC to coordinate disclosure is that companies typically do not pay bounties for vulnerabilities notified through this channel. This creates an incentive for security researchers to either notify bug-bounty vendors earlier than the rest of the affected ecosystem, or to exclude them from initial VINCE disclosures while claiming bounties in parallel.

\subsection{Open Source Disclosures}

Sharing embargoed vulnerability disclosures with open-source software maintainers is not always straightforward. Some teams expect issues to be raised in public on GitHub or other open platforms. 

Some projects have an established process for confidential disclosure; examples include the Rust and LLVM projects. However, GCC -- GNU's immensely popular C/C++ compiler -- does not at the time of writing advertise any method to send embargoed security reports.

We found that an effective way of getting through to such projects is via commercial open-source operating systems such as Red Hat. These organizations employ significant contributors to most critical open-source projects, and have an interest in ensuring that the open-source ecosystem is patched quickly. If a researcher sends a disclosure to, and requests assistance from, such a company, its employees can write patches privately for affected software and release them when the vulnerability is publicly disclosed.

One other key resource in ensuring pre-release preparation among the open-source community is the distros mailing list~\cite{distros_list}. This closed list is read by maintainers of most major Linux operating systems. It is willing to accept embargoes of up to 14 days in length, after which time the disclosures must be made public. This ticking clock is a helpful tool to nudge teams to install patches, or to pre-brief them on anticipated patch releases.

\subsection{CVEs}

CVEs (Common Vulnerabilities and Exposures) are universal identifiers that provide common references for discussing vulnerabilities~\cite{mitre_cve_2021}. We requested two CVEs for different variants of the Trojan Source attack on October 18, 2021. They were issued on the same day: CVE-2021-42574 and CVE-2021-42694.

CVEs are issued by CVE Number Authorities, or CNAs. Since many of our disclosure recipients were CNAs, we had initially hoped that one or more would issue a CVE for us. This did not happen. With hindsight, it is understandable that firms are reluctant to attach their brand to ecosystem-wide vulnerabilities.

Thankfully, MITRE -- the organization sponsoring the CVE program -- acts as the ``CNA of Last Resort''. We therefore requested CVEs from them directly against the Unicode Specification in the hope of motivating recipients to pay attention to our disclosures. We were surprised at the speed and simplicity of the process: one need only send a properly formatted email to a dedicated mailbox, and a CVE number is sent back shortly thereafter. MITRE does not appear to take a view on whether something is indeed a vulnerability. 

Our CVEs were helpful in motivating the disclosure process; we noticed a clear increase in attention after appending them to existing threads. This is slightly surprising given how easy CVEs are to get.

\subsection{Ecosystem Scanning}

During the coordinated disclosure process we were curious to learn if Trojan Source techniques were already being used in the wild. We initially tried out a variety of different ecosystem code search tools we found online, and although we found no signs of exploitation, we had low confidence that they indexed the relevant control characters well enough to identify the directionality attack.

We therefore partnered with GitHub to scan their backend for public repositories containing indicators of a directionality control character attack. This involved us providing a regex to GitHub, after which they performed the scan and sent us the results. We found some evidence of adversarial use of directionality control characters, specifically in crafting misleading smart contracts, but we did not find any use of the techniques described in our disclosures.

The Rust team also volunteered to scan their package index, crates.io, for the attacks, and did not find any indications of the attack pattern in their results.
\section{Public Release}

We publicly released information about the Trojan Source attacks on November 1, 2021 at 0:00 UTC. We now describe the release process, media coverage, and patches.

\subsection{Website}

\begin{figure}[t]
    \centering
    \includegraphics[width=\columnwidth]{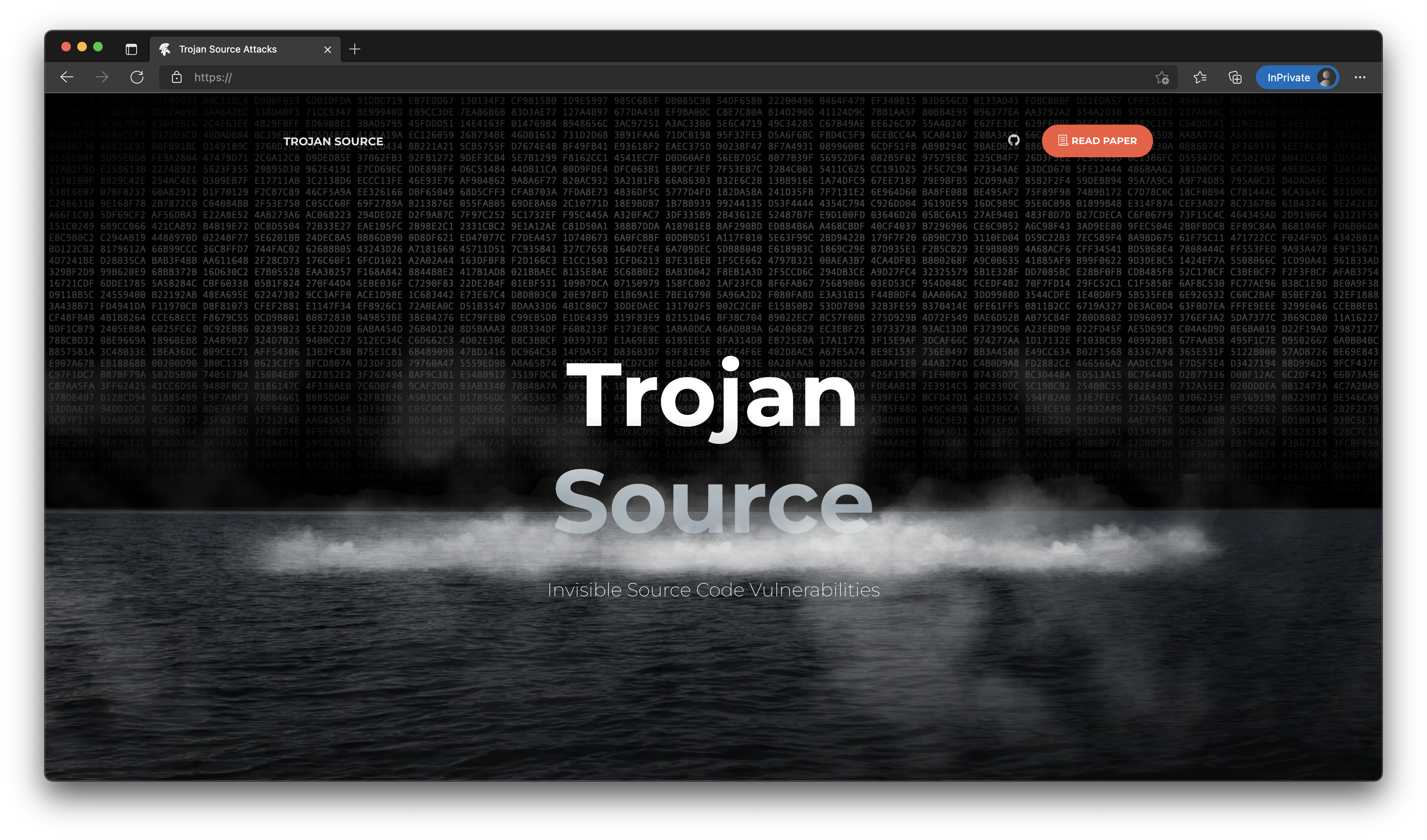}
    \caption{Website launched for public release of Trojan Source}
    \label{fig:website}
\end{figure}

Our primary public release method was a website\footnote{\href{https://trojansource.codes}{trojansource.codes}} which we launched at midnight UTC on November 1st. This website hosted a copy of our technical paper, a summary of the attack, and a link to proofs of concept published simultaneously on GitHub. A screenshot of the site can be seen in \Cref{fig:website}.

We tracked access to the site using a GDPR-compliant analytics tool~\cite{simpleanalytics} which logged 42,453 views by 38,888 unique visitors in the first 48 hours. At the time of writing, just under four months later, the site has been viewed 106,697 times by 92,762 unique users. During the same period, the GitHub repo has received 1,071 stars. The combination of a website, technical paper, and proofs-of-concept has become the standard way of disclosing vulnerabilities of systematic interest; we suspect that without the websites, significantly fewer people would read the technical papers.


\subsection{Press Coverage}

Two days prior to public release, we sent a draft of the Trojan Source paper to the authors of two security blogs and one tech news site. \textit{Krebs on Security} was the first to write about the attack~\cite{krebs_trojansource}, followed shortly by \textit{Schneier on Security}~\cite{schneier_trojansource}. Press coverage followed from The Register~\cite{register_trojansource}, Gizmodo~\cite{gizmodo_trojansource}, ZDNet~\cite{zdnet_trojansource}, Computer Weekly~\cite{computerweekly_trojansource}, Bleeping Computer~\cite{bleepingcomputer_trojansource}, LWN~\cite{lwn_trojansource}, and many others.

We also wrote a post linking to the website and paper on our laboratory's blog~\cite{lbt_trojansource} and tweeted it. Based on web referrers logged by our website analytics, Twitter was the most common discovery path followed by our blog post. Eventually, YouTube, Google, and GitHub joined the list of top referrers.

We were later contacted by two computer security podcasts -- DevNews~\cite{devnews_trojansource} and Cyberwire~\cite{cyberwire_trojansource} --  inviting us to discuss the work on their shows, which we did.

\subsection{Patches}

The Trojan Source attack can be mitigated at multiple stages in the software development pipeline including compilers/\allowbreak{}interpreters, code editors, and code repository web front ends. To simplify discussion, we may refer to the first of these as `the language' and the last two as `the editor'. Static code analysis tools can also play a role in mitigation. 

The fact that the attack can be blocked by either the language or the editor opens up the possibility of blame shifting. A language team that can't be bothered to patch can blame the editor, while the maintainers of an editor can similarly claim that vulnerable languages should be fixed instead.

In response to our disclosures, a wide array of software was patched in parallel with the public release of the attack methodology. In the following sections, we describe each public patch.

\subsubsection{Code Repositories}

\begin{figure}[t]
    \centering
    \includegraphics[width=\columnwidth]{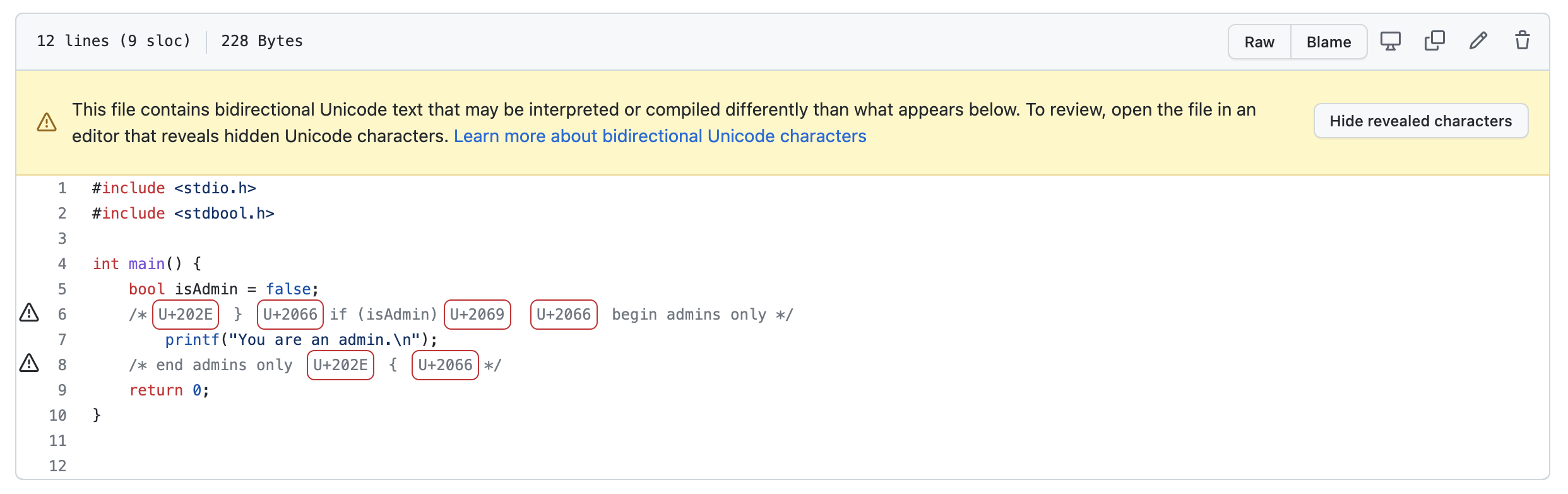}
    \caption{Trojan Source mitigations patched in the GitHub web UI}
    \label{fig:github}
\end{figure}

Three code repositories released patches to defend against Trojan Source attacks. The most prominent, GitHub, updated their web-based UI with mitigations and published a security advisory~\cite{github_advisory}. Their mitigations draw attention to bidirectional overrides by displaying a warning banner, a link to guidance, a warning symbol on the affected line, and optionally a visualization of the bidi character code points. No defenses appear to have been deployed, though, for the homoglyph and invisible-character variants of the attack.

Bitbucket, a web-based code repository produced by Atlassian, also released an advisory and deployed patches~\cite{atlassian_advisory}. Bitbucket now displays directionality control characters as Unicode code points by default; it does not, however, display any other warning messages. Nor does it have any defenses for the homoglyph and invisible-character variants.

GitLab, another web-based code repository, also published an advisory and released a patch~\cite{gitlab_advisory}. Their defense displays all bidi characters as the \ucr\ symbol with a red underline. GitLab is the one repo front-end to provide a defense against homoglyph attacks: it highlights suspect homoglyphs in red. Here too, invisible characters remain invisible.

\subsubsection{Code Editors}

\begin{figure}[t]
    \centering
    \includegraphics[width=\columnwidth]{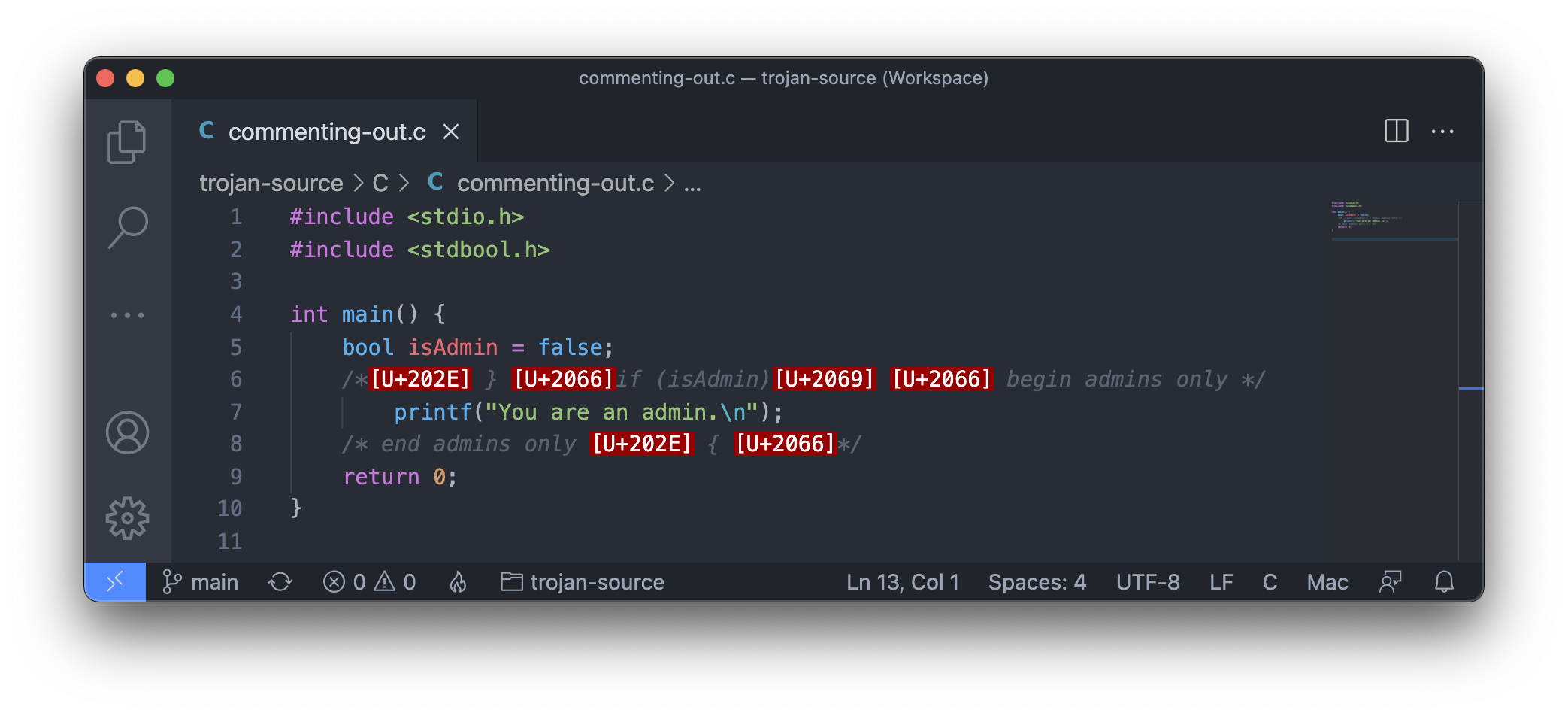}
    \caption{Trojan Source mitigations patched in Visual Studio Code}
    \label{fig:vscode}
\end{figure}

Four code editors also deployed patches to defend against Trojan Source attacks. Visual Studio Code patched the UI~\cite{vscode_advisory} so that bidi control characters are rendered as code points and highlighted in red as in \Cref{fig:vscode}. Suspect homoglyphs are also highlighted by rendering a yellow box around them.

Another code editor that released a patch was Emacs, which now highlights any suspected adversarial use of bidirectional control characters~\cite{emacs_patch}.

Two other code editors, Visual Studio and Sublime Text, now simply ignore  directionality control characters in source code, which could be considered a partial mitigation.

\subsubsection{Compilers}


We argued in our technical paper that the most robust place to defend against Trojan Source attacks is in programming language specifications, as requirements there specified are guaranteed to be implemented by language-compliant compilers and interpreters. However, not all languages have formal specifications, and even for those that do it may be prudent to have interim defenses in the form of compiler errors or warnings, as specification changes can take a long time to be agreed and implemented.

\begin{figure}[t]
    \centering
    \vspace{1em}
    \includegraphics[width=\columnwidth]{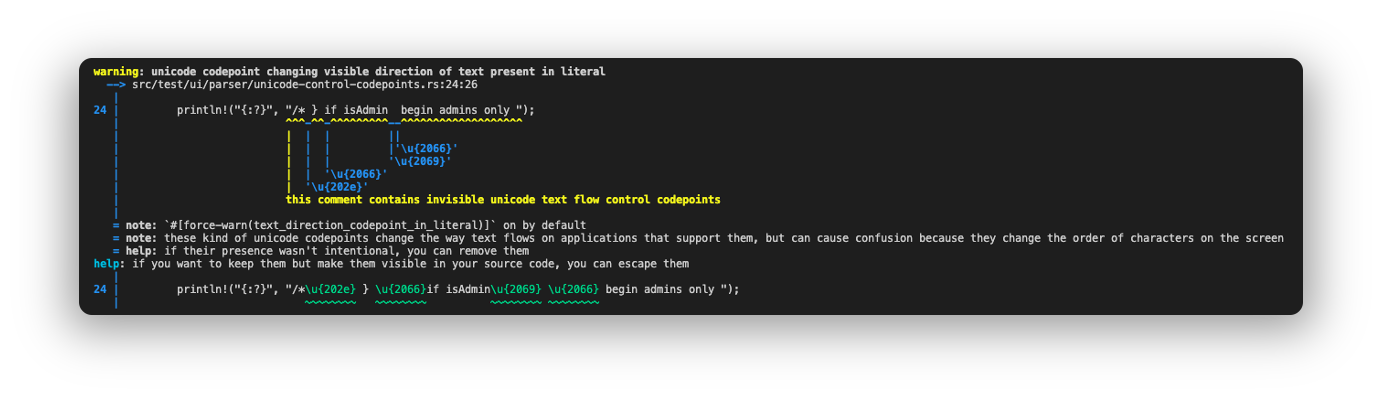}
    \caption{Trojan Source warning in Rust compiler}
    \label{fig:rust}
\end{figure}

Of the compiler teams that we contacted, the Rust team was both eager to implement defenses and one of the most helpful teams to work with. Rust published a security advisory and compiler update in parallel with our public release of the attack~\cite{rust_advisory}. The Rust compiler patch included a default-enabled warning identifying directionality control characters, which we show in \Cref{fig:rust}. Interestingly, the Rust compiler already had mitigations to warn against the homoglyph variant of the attack.

GCC, a common C and C++ compiler produced by GNU, took a similar approach to Rust and later launched a default-enabled warning \texttt{-Wbidi-chars} that sounds an alarm for suspected Trojan Source attacks~\cite{gcc_wbidichars}.

Julia, a high-performance scientific language, followed the recommendation in our technical paper and disallowed unterminated bidirectional control characters in comments and string literals~\cite{julia_release}. As Julia was not a recipient on our embargoed disclosure list, their action illustrates the benefit of public disclosure.

LLVM, the system underlying the alternate common C and C++ compiler \texttt{clang}, took a slightly different approach: rather than adding errors in the compiler itself, the project maintainers added checks to the accompanying linter \texttt{clang\-tidy}. These checks provide alerts both for directionality and homoglyph attacks, and were announced in a dedicated security advisory~\cite{llvm_advisory}. This is a helpful partial mitigation, but it will only benefit users that run \texttt{clang-tidy} in addition to \texttt{clang}.

Not all compiler teams agreed with implementing mitigations, however. Java was a key outlier in their response to Trojan Source attacks. Oracle, the maintainers of Java, provided the following response to our disclosure:

\begin{displayquote}
This report is not about a bug in Java as such, but rather about a code constructed in such a way it may confuse reviewers, to allow exploits to be sneaked into other projects. While we would encourage review and similar tools to employ heuristics to detect suspicious constructs, whether or not presented in this bug, the Java team does not provide review tools, and hence the ability to help with this problem from the Java platform is limited.
\end{displayquote}

The issue was subsequently closed as ``Not a Bug''. Other languages, such as Node.js, also considered these attacks not a bug, but often with additional justification. Node.js, for example, calls out the challenges in alerting for errors of this kind in interpreted languages. Unlike statically compiled languages, the error will not be detected until runtime and is therefore more likely to cause problems in the case of false positives. Due to this, Node.js recommends that their users use code scanning tools to detect these attacks.

Other compiler teams were less committal. Python indicated that Trojan Source attacks may be tackled in future versions of the language~\cite{python_response}, and we did not receive any commitments from Go or C\#.

\subsection{Conference Submissions}

During the coordinated disclosure period, we submitted a paper describing Trojan Source attacks to the 43rd IEEE Symposium on
Security and Privacy. However, the date for public release that we had already negotiated amongst disclosure recipients and committed to fell during the review cycle for the conference submission. According to conference submission rules, ``authors may choose to give talks about their work, post a preprint of the paper to an archival repository such as arXiv, and disclose security vulnerabilities to vendors. Authors should refrain from widely advertising their results, but in special circumstances they should contact the PC chairs to discuss exceptions''~\cite{oakland_cfp}.

Although the conference rules permit disclosure of security vulnerabilities, we sought written permission from the program committee chairs to publish information about the attack when it was publicly released, thus ensuring compliance with publicity restrictions. This permission was granted; one of the conference chairs confirmed that since our public release was scheduled following the rebuttal period it should not interfere with the review process.

To our surprise, the paper was rejected despite initial reviews that were much more positive than those for another paper that was accepted at the same conference. The reviewers gave breaking anonymity via publicity as one of the reasons for rejection, and also referenced URLs of online discussions in their rejection showing that they had personally read the coverage. We appealed the rejection to the program committee chairs, citing their prior written approval to release the paper on the specified date, to which we received a reply that their approval had been a mistake.

\subsection{Unicode Working Group}

Following the publication of Trojan Source, the Unicode Consortium announced the formation of a working group to address the issues raised by Trojan Source attacks~\cite{unicode_avoiding_spoofing}. We have been contacted by the working group with various questions, and expect that a future version of the Unicode Specification will provide guidance to help mitigate Trojan Source attacks.

Indeed, we suspect that the long-term fix for Trojan Source attacks will be driven from changes to the Unicode specification. Unicode already provides security guidance for some aspects of source code, such as the characters that should be considered permissible for identifiers in code. Adding similar guidance for directionality override characters and documenting methods to identify homoglyph and invisible character-based attacks should go a long way in solving the issue. Guidance in the Unicode Specification is likely to be adopted downstream by language specifications, and this in turn is likely to be implemented by maintainers of compilers and interpreters. However this is the slow boat, and may take several years to work its way round the world. 

In the meantime, many languages remain vulnerable, so mitigations in editors and code-scanning tools will remain essential for any critical project to which adversarial contributions are only blocked by human code review. In an ideal world, such projects would use a defense-in-depth strategy: vulnerabilities that cross domain boundaries along a tool chain or a supply chain, such as Trojan Source, should ideally be mitigated at more than one point in the chain.

\section{Discussion}

\subsection{Historical background}

Vulnerability disclosure has been a topic of interest for twenty years now. In 2002 Jean Camp proposed vulnerability markets, which emerged shortly afterwards~\cite{Camp2002}. 2004 saw a debate between Eric Rescorla, who argued on the basis of data from 1988-2003 that disclosing vulnerabilities publicly rather than privately did not obviously lead to more rapid vulnerability depletion~\cite{Rescorla2004}, and Ashish Arora who argued that the improved incentive for bug fixing tipped the balance in favour of public disclosure, albeit after a delay~\cite{ATX2004}. The following year, Andy Ozment published a paper with data on the likelihood of vulnerability rediscovery, showing that the rate of vulnerability discovery in OpenBSD was declining over time~\cite{Ozment2005}; at the same workshop, Ashish Arora and colleagues had data showing that disclosure caused firms to patch significantly more quickly~\cite{AKTY2005}. The following year saw not just multiple models of how patch management might work in theory, but also a paper by Michael Sutton and Frank Nagle of iDefense, one of the first firms to operate a vulnerability market, reporting how it worked in practice~\cite{SN2006}.

By this time the argument in favour of responsible disclosure had been won in the core of the tech industry, against both the open-disclosure radicals (who favoured releasing all bugs anonymously in public on lists such as bugtraq without giving firms a chance to patch them), and the traditionalists of the defense establishment and corporate legal departments (who wanted all disclosure to be suppressed by the civil or even criminal law). This consensus has not propagated everywhere; as late as 2013, Volkswagen sued researchers at the universities of Birmingham and Nijmegen after they responsibly disclosed a vulnerability in the car company's remote key entry system; but they lost the resulting court case~\cite{Prigg2015}. 

The patching ecosystem became more adversarial after 2013 when the Stuxnet worm alerted governments to the potential use of vulnerabilities in cyber-weapons, and firms emerged that bought them for sale to government agencies and to cyber-arms manufacturers that work for governments. Competition from these exploit acquisition firms has driven up the prices of zero-click vulnerabilities in popular platforms such as Android and iOS into six and even seven figures, compared with the four-to-five figures reported in 2006. This story is told by Nicole Perlroth~\cite{Perlroth2020}.

Complexity has also increased thanks to the depth and breadth of modern supply chains. A vulnerability in a widely-used platform such as Linux, or a widely-used library such as OpenSSL, can force thousands of firms to scramble to patch their products. Kiran Sridhar and colleagues analyse the metadata of 434k emails sent through CERT/CC since 1993 about 46k vulnerabilities to devise the patterns; vulnerabilities further up the supply chain take longer to coordinate, and those affecting more vendors require more communication. CERT/CC is also more likely to coordinate things where there is a public exploit, or where there is no capable vendor willing to lead the remediation effort.

\subsection{Improving Disclosure Incentives}

There is a direct financial benefit to all security researchers, whether industrial, academic, or hobbyist, who submit vulnerabilities to bug bounty programs. But not all vendors offer bounties, and those that do are often spread across multiple platforms. They also typically limit further disclosure while the issue is being repaired, which can be in tension with coordinated disclosure, where the goal is to inform many different entities.

In our case, we sent our disclosure to several bug bounty programs before discovering the centralized CERT/CC process. This process can minimize time and maximize impact: researchers need only disclose the vulnerability once, they receive staff assistance, and they can answer follow-up questions in a single interface. However, a disclosure via CERT/CC is, to the best of our knowledge, not eligible for the bug bounties offered by any major company.

Current programs thus provide the wrong incentives for supply chain or broad-impact vulnerabilities. A rational, strategic actor will submit many reports to separate bug bounty programs, rather than engaging in widespread coordinated disclosure. This is bad for everyone. Organizations without programs are less likely to be able to build patches during an embargo period, while security researchers will spend more time on communications and less on research. Even companies that do offer bounties may be negatively impacted if they consume software that goes unpatched.

We therefore recommend that all bug bounty programs should include coordinated disclosures in their scope. Ideally, they would not only reward, but actively encourage, the disclosure of cross-organization vulnerabilities via shared channels or tools such as CERT/CC. This would re-align incentives for disclosure, make better use of existing tools, and enhance the technical security posture of the software ecosystem.

\subsection{Academic Publication}

Most top computer science conferences use anonymous peer review~\cite{9833581}, though there is some variation in procedure; further, some information about author identity will inevitably leak via submitted artefacts, the citation of prior work, and from program committee members having seen talks about work in progress~\cite{10.1145/3208157}. In the context of security research, the expectation that the burden of anonymity falls mostly on authors impedes the effectiveness of vulnerability disclosure.

Public release is the key component in vulnerability disclosure: the countdown to disclosure pushes firms to repair their software quickly. Advertising the vulnerability helps nudge users to install patches or other mitigations, while also helping to flush out other software that is impacted but was not initially patched. Unfortunately, advertisement that is effective is likely to break anonymity. 

Is it possible to achieve the ecosystem benefits of disclosure while maintaining the scientific benefits of anonymous peer review? We believe that the answer is yes. Rather than place the onus on authors to ensure that reviewers don't discover their identities online, it should be the responsibility of reviewers to not seek out information about the identities of authors. Reviewers should be asked to avoid searching for online coverage of material in the papers they are reviewing, and to disregard anything they think they recall. Authors may be asked to anonymize their conference paper submissions as far as is reasonably practical, but program committees should refrain from gold-plating this requirement, and the responsibility of keeping reviews anonymous should be shared sensibly with the reviewers.

\subsection{Machine-Learning Disclosures}
\label{sec:ml_disclosures}

Our work on Trojan Source Attacks was largely based on prior work (with a different set of authors) in using Unicode perturbations to attack machine-learning NLP models~\cite{boucher_2022_badchars}. That work found that nearly all text-based machine-learning models are vulnerable to adversarial examples crafted using Unicode techniques such as directionality overrides, homoglyph substitutions, and the injection of invisible characters. These adversarial examples are crafted using visually imperceptible perturbations; that is, an adversarial example looks like a benign piece of text when displayed to a user, but its encoding causes machine-learning models to output either low-quality or adversarially targeted outputs during inference.

The authors of this prior work notified the companies and organizations producing the models they found they could break. They also proposed defenses,  ranging from deterministic pre-processing of inputs to using optical-character recognition to map visual renderings to consistent Unicode representations. Rather than submitting these vulnerabilities to bug bounty programs, they just notified contacts at the affected companies. It did not seem at the time that machine-learning pipeline vulnerabilities would be considered in-scope for bug bounty programs.

It has been over a year since contact was made with the affected companies, and virtually no changes have been made. The one exception is Google, which appears to have deployed an update to its Google Translate model that makes it robust against homoglyph substitutions and invisible character injections -- although it is still vulnerable to bidi control characters. Such vulnerabilities enable a range of attacks on systems that process textual input. Hate speech can be hidden from filters, search can be misdirected, and text crafted so that automated translations are wrong in targeted ways. More than a year after disclosure, the vulnerable systems are still used at scale for a wide range of societally important tasks.

This points to a larger problem with bugs that cross the domain boundary of two communities -- here, the security and NLP communities. Each community, even within a company, can be tempted to blame the other, and expect someone else to fix the problem. This is an obvious externality, and although we might normally expect that externalities can be dealt with by firms that are large enough to internalize them, this is largely not happening here. If even single companies cannot identify and handle subtle vulnerabilities in machine-learning pipelines, this does not bode well for wider ecosystems. 

There are many possible reasons why ML/NLP systems might not get patched as quickly, or at all. First, patches that involve retraining a large model can take time and cost money. Second, the culture of C programmers is very different from that of data scientists; people who build operating systems expect that they'll have to ship patches quickly. Third, there are different expectations of dependability. Fourth, these attitudes are reflected in the press; there was much greater coverage of the Trojan Source vulnerability on code than of the very similar Bad Characters attack on NLP. And finally there's a matter of maturity, of both the technology and the market.

As traditional attacks like buffer overflows are supplanted by more modern attacks such as adversarial examples~\cite{szegedy2013intriguing,goodfellow2014explaining,boucher_2022_badchars}, patch management will get still harder. There is often disagreement about what is considered a vulnerability, and it is unclear whether mitigations within organizations should be driven by traditional security teams or by machine-learning teams. As we depend more and more on machine-learning components, we will have to establish shared definitions of vulnerabilities along with norms for defense ownership. We will also have to embed security expertise within machine-learning teams, and probably develop new ways of engineering security end-to-end for systems that contain machine-learning components.

\subsection{Vulnerability Types}

Vulnerability type is likely to affect the coordinated disclosure process. In \Cref{sec:outsourced_platforms}, we noted that novel vulnerability patterns are less likely to garner engagement during the report recipient triage process, and in \Cref{sec:ml_disclosures} we noted that there is often disagreement over what is considered a vulnerability in ML systems. 
We believe that the experience of disclosing the single vulnerability in this study is of wider interest, and that the literature will benefit from additional studies published with experiences related to other vulnerability patterns.
\section{Conclusion}

A vulnerability that affected almost all programming languages and most editors, and which can still be used to insert malicious code into any project that uses a combination of an unsafe programming language and an unsafe editor, has provided an interesting natural experiment in coordinated disclosure. Some versions of the vulnerability have so far been fixed in most editors and some programming languages. An analogous vulnerability in NLP-based machine learning systems has not evinced a similar response, having received only a partial fix in the offerings of a single service provider. In this paper, we have described the tools, timeline, and stakeholders involved in the coordinated disclosure process, discussed their response and analyzed the outcomes. We have also made some recommendations to improve the system.

Vulnerability research tends to focus on technical findings and on the actual repairs needed to software systems, and even in the security-economics community, most attention has been given to the post-release period of disclosures. Yet there is real potential for practical improvement in the disclosure process from research on the often cloaked, pre-public phase of vulnerability disclosure, and on the incentives facing the various actors in the modern world of bug bounties and outsourced platforms. This will become ever more important as more and more disclosures are coordinated across multiple actors in complex supply chains.

\printbibliography

@article{boucher_trojansource_2021,
    title = {Trojan {Source}: {Invisible} {Vulnerabilities}},
    author = {Nicholas Boucher and Ross Anderson},
    year = {2021},
    journal = {Preprint},
    eprint = {2111.00169},
    archivePrefix = {arXiv},
    primaryClass = {cs.CR},
    url = {https://arxiv.org/abs/2111.00169}
}

@misc{Camp2002,
    title={Marketplace Incentives to Prevent Piracy: An Incentive for Security?},
    author={L. Jean Camp},
    organization={{WEIS}},
    year={2002}
}

@misc{Rescorla2004,
    title={Is finding security holes a good idea?},
    author={Eric Rescorla},
    organization={{WEIS}},
    year={2004},
    url={https://ieeexplore.ieee.org/document/1392694}
}

@misc{ATX2004,
    title={Optimal Policy for Software Vulnerability Disclosure},
    author={Ashish Arora and Rahul Telang and Hao Xu},
    organization={{WEIS}},
    year={2004},
    url={https://www.jstor.org/stable/20122417}
}

@misc{Ozment2005,
    title={The Likelihood of Vulnerability Rediscovery and the Social Utility of Vulnerability Hunting},
    author={Andy Ozment},
    organization={{WEIS}},
    year={2005}
}

@misc{AKTY2005,
    title={Ashish Arora, Ramayya Krishnan, Rahul Telang, Yubao Yang},
    author={Ashish Arora and Ramayya Krishnan and Rahul Telang and Yubao Yang},
    organization={{WEIS}},
    year={2005}
}

@misc{SN2006,
    title={Emerging Economic Models for Vulnerability Research},
    author={Michael Sutton and Frank Nagle},
    organization={{WEIS}},
    year={2006},
    url={https://econinfosec.org/archive/weis2006/docs/17.pdf}
}

@misc{Prigg2015,
    title={Hackers reveal flaw in over 100 cars kept secret by Volkswagen for TWO YEARS: Bug can be used to unlock everything from a Kia to a Lamborghini},
    author={Mark Prigg},
    year={2015},
    organization={{Daily Mail}}
}

@book{Perlroth2020,
    address = {New York},
    author = {Nicole Perlroth},
    publisher = {Bloomsbury Publishing},
    title = {This is How They Tell Me the World Ends},
    year = {2020}
 }

@misc{log4j_2021,
 title={{Apache Log4j Security Vulnerabilities}}, url={https://logging.apache.org/log4j/2.x/security.html},
 organization={{The Apache Software Foundation}},
 author={{The Apache Software Foundation}},
 year={2021},
 month=dec,
 day={28}
 }

@article{solarwinds_2021,
author = {S. Peisert and B. Schneier and H. Okhravi and F. Massacci and T. Benzel and C. Landwehr and M. Mannan and J. Mirkovic and A. Prakash and J. Michael},
journal = {IEEE Security \& Privacy},
title = {Perspectives on the SolarWinds Incident},
year = {2021},
volume = {19},
number = {02},
issn = {1558-4046},
pages = {7-13},
doi = {10.1109/MSEC.2021.3051235},
publisher = {IEEE Computer Society},
address = {Los Alamitos, CA, USA},
month = mar
}

@misc{biden_2021,
	title = {Executive {Order} on {Improving} the {Nation}'s {Cybersecurity}},
	url = {https://www.whitehouse.gov/briefing-room/presidential-actions/2021/05/12/executive-order-on-improving-the-nations-cybersecurity},
	language = {en-US},
	urldate = {2021-07-05},
	author = {Joseph Biden},
	month = may,
	year = {2021},
	note = {{Executive} {Order} 14028}
}

@misc{unicode_2021,
    author = {{The Unicode Consortium}},
	title = {The {Unicode} {Standard}, {Version} 14.0},
	url = {https://www.unicode.org/versions/Unicode14.0.0},
	language = {en},
	organization = {The Unicode Consortium},
	month = sep,
	year = {2021}
}

@techreport{unicode_bidi_2021,
    author = {{The Unicode Consortium}},
	title = {Unicode {Bidirectional} {Algorithm}},
	url = {https://www.unicode.org/reports/tr9/tr9-44.html},
	language = {en},
	number = {Unicode Technical Report \#9},
	institution = {The Unicode Consortium},
	day = 27,
	month = aug,
	year = {2021}
}

@inproceedings{boucher_2022_badchars,
    title = {Bad {Characters}: {Imperceptible} {NLP} {Attacks}},
    author = {Nicholas Boucher and Ilia Shumailov and Ross Anderson and Nicolas Papernot},
    booktitle = {43rd IEEE Symposium on Security and Privacy},
    year = {2022},
    organization = {IEEE},
    url={https://ieeexplore.ieee.org/document/9833641}
}

@misc{certcc,
	title = {{CERT Coordination Center}},
	url = {https://www.kb.cert.org},
	urldate = {2021-10-29},
	author = {{Carnegie Mellon University Software Engineering Institute}}
}

@misc{distros_list,
     author={{Openwall Project}},
     title={Operating System Distribution Security Contact lists},
     url={https://oss-security.openwall.org/wiki/mailing-lists/distros},
     year={2021},
     month=sep
 }

@misc{krebs_trojansource,
    author={Brian Krebs},
    title={{‘Trojan Source’ Bug Threatens the Security of All Code}},
    url={https://krebsonsecurity.com/2021/11/trojan-source-bug-threatens-the-security-of-all-code},
    organization = {{Krebs on Security}},
    year={2021},
    month=nov,
    day={1}
}

@misc{schneier_trojansource,
    author={Bruce Schneier},
    title={{Hiding Vulnerabilities in Source Code}},
    url={https://www.schneier.com/blog/archives/2021/11/hiding-vulnerabilities-in-source-code.html},
    organization = {{Schneier on Security}},
    year={2021},
    month=nov,
    day={1}
}

@misc{register_trojansource,
    author={Gareth Corfield},
    title={{Trojan Source attack: Code that says one thing to humans tells your compiler something very different, warn academics}},
    url={https://www.theregister.com/2021/11/01/trojan_source_language_reversal_unicode},
    organization = {{The Register}},
    year={2021},
    month=nov,
    day={1}
}

@misc{gizmodo_trojansource,
    author={Lucas Ropek},
    title={{Pretty Much All Computer Code Can Be Hijacked by Newly Discovered 'Trojan Source' Exploit}},
    url={https://gizmodo.com/pretty-much-all-computer-code-can-be-hijacked-by-newly-1847974191},
    organization = {{Gizmodo}},
    year={2021},
    month=nov,
    day={1}
}

@misc{zdnet_trojansource,
    author={Liam Tung},
    title={{Programming languages: This sneaky trick could allow attackers to hide 'invisible' vulnerabilities in code}},
    url={https://www.zdnet.com/article/this-sneaky-trick-could-allow-attackers-to-hide-invisible-vulnerabilities-in-code},
    organization = {{ZDNet}},
    year={2021},
    month=nov,
    day={1}
}

@misc{computerweekly_trojansource,
    author={Bill Goodwin},
    title={{Businesses and governments urged to take action over Trojan Source supply chain attacks}},
    url={https://www.computerweekly.com/news/252508879/Businesses-and-governments-urged-to-take-action-over-Trojan-Source-supply-chain-attacks},
    organization = {{Computer Weekly}},
    year={2021},
    month=nov,
    day={1}
}

@misc{bleepingcomputer_trojansource,
    author={Ionut Ilascu},
    title={{'Trojan Source' attack method can hide bugs into open-source code}},
    url={https://www.bleepingcomputer.com/news/security/trojan-source-attack-method-can-hide-bugs-into-open-source-code},
    organization = {{Bleeping Computer}},
    year={2021},
    month=nov,
    day={1}
}

@misc{lwn_trojansource,
    author={Jake Edge},
    title={{Trojan Source: tricks (no treats) with Unicode}},
    url={https://lwn.net/Articles/874951},
    organization = {{LWN}},
    year={2021},
    month=nov,
    day={3}
}

@misc{hackerone,
    author={{HackerOne}},
    title={{Bug Bounty Platform}},
    url={https://www.hackerone.com/product/bug-bounty-platform},
    organization = {{HackerOne}},
    year={2022}
}

@misc{bugcrowd,
    author={{BugCrowd}},
    title={{Managed Bug Bounty}},
    url={https://www.bugcrowd.com/products/bug-bounty},
    organization = {{BugCrowd}},
    year={2022}
}

@misc{simpleanalytics,
    author={{Simple Analytics}},
    title={{The privacy-first Google Analytics alternative}},
    url={https://simpleanalytics.com},
    organization = {{Simple Analytics}},
    year={2022}
}

@misc{lbt_trojansource,
    author={Ross Anderson and Nicholas Boucher},
    title={{Trojan Source: Invisible Vulnerabilities}},
    url={https://www.lightbluetouchpaper.org/2021/11/01/trojan-source-invisible-vulnerabilities},
    organization = {{Light Blue Touchpaper}},
    year={2021},
    month=nov,
    day={1}
}

@misc{devnews_trojansource,
    author={Saron Yitbarek and Josh Puetz},
    title={{No More Contacting Employees Off Hours in Portugal, Trojan Source Attacks, Another Apple Settlement, \& more on DevNews!}},
    url={https://dev.to/devteam/no-more-contacting-employees-off-hours-in-portugal-trojan-source-attacks-another-apple-settlement-more-on-devnews-59i1},
    organization = {{DEV}},
    year={2021},
    month=nov,
    day={18}
}

@misc{cyberwire_trojansource,
    author={Dave Bittner},
    title={{Trojan Source--a threat to the software supply chain. Ransomware goes to influence operations school. Triple extortion? Criminal target selection.}},
    url={https://thecyberwire.com/podcasts/daily-podcast/1451/notes},
    organization = {Cyberwire},
    year={2021},
    month=nov,
    day={2}
}

@misc{github_advisory,
	title = {Warning about bidirectional Unicode text},
	url = {https://github.blog/changelog/2021-10-31-warning-about-bidirectional-unicode-text},
	urldate = {2022-1-30},
	organization = {GitHub},
	author = {{GitHub}},
	day = 31,
	month = oct,
	year = {2021}
}

@misc{atlassian_advisory,
	title = {{Multiple Products Security Advisory - Unrendered unicode bidirectional override characters - CVE-2021-42574}},
	url = {https://confluence.atlassian.com/security/multiple-products-security-advisory-unrendered-unicode-bidirectional-override-characters-cve-2021-42574-1086419475.html},
	urldate = {2022-1-31},
	organization = {Atlassian},
	author = {{Atlassian}},
	day = 1,
	month = nov,
	year = {2021}
}

@misc{gitlab_advisory,
	title = {{GitLab Security Release: 14.4.1, 14.3.4, and 14.2.6}},
	url = {https://about.gitlab.com/releases/2021/10/28/security-release-gitlab-14-4-1-released/},
	urldate = {2022-1-31},
	publisher = {GitLab},
	author = {{GitLab}},
	day = 28,
	month = oct,
	year = {2021}
}

@misc{vscode_advisory,
	title = {{Visual Studio Code: October 2021 (version 1.62)}},
	url = {https://code.visualstudio.com/updates/v1_62},
	urldate = {2022-1-31},
	publisher = {Microsoft},
	author = {{Microsoft}},
	month = oct,
	year = {2021}
}

@misc{emacs_patch,
	title = {{Better detection of potentially malicious bidi text}},
	url = {https://git.savannah.gnu.org/cgit/emacs.git/commit/?id=b96855310efed13e0db1403759b686b9bc3e7490},
	urldate = {2022-1-31},
	organization = {GNU},
	author = {Eli Zaretskii},
	day = 4,
	month = nov,
	year = {2021}
}

@misc{unicode_avoiding_spoofing,
	title = {{Avoiding Source Code Spoofing}},
	url = {https://www.unicode.org/L2/L2022/22007r2-avoiding-spoof.pdf},
	organization = {Unicode},
	author = {Mark Davis and Robin Leroy and Peter Constable and Markus Scherer},
	day = 25,
	month = jan,
	year = {2022}
}

@misc{rust_advisory,
	title = {{Security advisory for rustc (CVE-2021-42574)}},
	url = {https://blog.rust-lang.org/2021/11/01/cve-2021-42574.html},
	urldate = {2022-1-31},
	publisher = {{The Rust Security Response WG}},
	author = {{The Rust Security Response WG}},
	day = 1,
	month = nov,
	year = {2021}
}

@misc{gcc_wbidichars,
	title = {{GCC: Warning Options}},
	url = {https://gcc.gnu.org/onlinedocs/gcc/Warning-Options.html},
	urldate = {2022-1-31},
	publisher = {GNU},
	author = {\vspace{0mm}GNU},
	day = 12,
	month = jan,
	year = {2022}
}

@misc{llvm_advisory,
	title = {{New passes in clang-tidy to detect (some) Trojan Source}},
	url = {https://blog.llvm.org/posts/2022-01-12-trojan-source},
	urldate = {2022-1-31},
	publisher = {LLVM},
	author = {LLVM},
	day = 12,
	month = jan,
	year = {2021}
}

@misc{oakland_cfp,
	title = {{Call For Papers}},
	url = {https://www.ieee-security.org/TC/SP2022/cfpapers.html},
	organization = {{43rd IEEE Symposium on
Security and Privacy}},
	author = {{IEEE}},
	year = {2021}
}

@article{10.1145/3208157,
    author = {Goues, C. Le and Brun, Y. and Apel, S. and Berger, E. and Khurshid, S. and Smaragdakis, Y.},
    title = {Effectiveness of Anonymization in Double-Blind Review},
    year = {2018},
    issue_date = {June 2018},
    organization = {Association for Computing Machinery},
    address = {New York, NY, USA},
    volume = {61},
    number = {6},
    issn = {0001-0782},
    url = {https://doi.org/10.1145/3208157},
    doi = {10.1145/3208157},
    journal = {Commun. ACM},
    month = may,
    pages = {30–33},
    numpages = {4}
}

@misc{julia_release,
	title = {{Julia v1.7 Release Notes}},
	url = {https://docs.julialang.org/en/v1.7/NEWS/#Language-changes},
	author = {{Julia Language Project Contributors}},
	day = 11,
	month = nov,
	year = {2021}
}

@misc{python_response,
    title = {{PEP 672 -- Unicode-related Security Considerations for Python}},
    url = {https://www.python.org/dev/peps/pep-0672},
    author = {Petr Viktorin},
    organization = {Python Software Foundation},
    day = 1,
    month = nov,
    year = {2021}
}

@article{szegedy2013intriguing,
  title={Intriguing properties of neural networks},
  author={Szegedy, Christian and Zaremba, Wojciech and Sutskever, Ilya and Bruna, Joan and Erhan, Dumitru and Goodfellow, Ian and Fergus, Rob},
  journal={arXiv preprint arXiv:1312.6199},
  year={2013}
}

@article{goodfellow2014explaining,
  title={Explaining and harnessing adversarial examples},
  author={Goodfellow, Ian J and Shlens, Jonathon and Szegedy, Christian},
  journal={International Conference on Learning Representations (ICLR)},
  year={2015}
}

@misc{mitre_cve_2021,
	title = {About the {CVE} {Program}},
	author = {{MITRE}},
	url = {https://www.cve.org/About/Overview},
	urldate = {2021-10-29},
	month = oct,
	year = {2021}
}

@INPROCEEDINGS{9833581,  author={Soneji, Ananta and Kokulu, Faris Bugra and Rubio-Medrano, Carlos and Bao, Tiffany and Wang, Ruoyu and Shoshitaishvili, Yan and Doupé, Adam},  booktitle={2022 IEEE Symposium on Security and Privacy (SP)},   title={“{F}lawed, but like democracy we don’t have a better system”: {T}he {E}xperts’ {I}nsights on the {P}eer {R}eview {P}rocess of {E}valuating {S}ecurity {P}apers},   year={2022},  volume={},  number={},  pages={1845-1862},  doi={10.1109/SP46214.2022.9833581}}


\end{document}